\documentclass[aps,prl,reprint,amsmath,amssymb,superscriptaddress]{revtex4-1}
\usepackage{graphicx}
\usepackage{color}
\usepackage{comment}
\usepackage{hyperref}
\usepackage[normalem]{ulem}
\newcommand{\clr}{\color{red}}
\newcommand{\clb}{\color{blue}}

\newcommand{\pbsout}[1]{{\color{blue} \sout{#1}}}
\begin{document}

\title{Topological band and superconductivity in UTe$_2$}

\author{Tatsuya Shishidou}
\affiliation{Department of Physics, University of Wisconsin,
  Milwaukee, WI~53201, USA}

\author{Han Gyeol Suh}
\affiliation{Department of Physics, University of Wisconsin,
  Milwaukee, WI~53201, USA}

\author{P. M. R. Brydon}
\affiliation{Department of Physics and MacDiarmid Institute for
  Advanced Materials and Nanotechnology, University of Otago,
  P.O.~Box~56, Dunedin~9054, New Zealand}
  
 \author{Michael Weinert}
\affiliation{Department of Physics, University of Wisconsin,
  Milwaukee, WI~53201, USA}

\author{Daniel F. Agterberg}
\affiliation{Department of Physics, University of Wisconsin,
  Milwaukee, WI~53201, USA}


\begin{abstract}

UTe$_2$ has recently been found to be a likely spin-triplet superconductor that exhibits evidence
for chiral Majorana edge states.  A characteristic structural
  feature of UTe$_2$ is inversion-symmetry related pairs of U atoms,
 forming rungs
of ladders. Here we show how  each rung's two 
sublattice degrees of freedom play a key role in
understanding the electronic structure and the origin of superconductivity.  In particular, we show
that DFT+$U$ calculations generically reveal a topological band that originates
from a band inversion associated with $5f$ electrons residing on these rung sublattice degrees of freedom. Furthermore, we show that a previously identified strong ferromagnetic interaction 
within a U-U rung 
leads to a pseudospin-triplet superconducting state that can account for a non-zero polar Kerr angle, observed magnetic field-temperature phase diagrams, and nodal Weyl
fermions.  Our analysis may also
  be relevant for other U-based superconductors.

\end{abstract}

\maketitle

UTe$_2$ \cite{Ran:2019} is poised to become a paradigmatic superconductor exhibiting unconventional
behavior: Superconductivity survives to much higher magnetic fields than expected \cite{Aoki:2019-1,Ran:2019,Knafo:2019,Miyake:2019,Imajo:2019,Mineev:2020,Ran:2020,Knebel:2020,Niu:2020,Niu:2020} and shows a highly unusual re-entrant field induced superconductivity \cite{Ran:2019-2}. Furthermore,  there is evidence for ferromagnetic fluctuations \cite{Tokunaga:2019,Sundar:2019}, odd-parity superconductivity \cite{Nakamine:2019}, multiple superconducting phases \cite{Braithwaite:2019,Hayes:2020,Thomas:2020,Machida:2020,Kittaka:2020}, spontaneous broken time-reversal symmetry \cite{Hayes:2020}, and chiral Majorana edge and surface states \cite{Bae:2019,Jiao:2020}, the nature of which are not yet understood. 

\begin{figure}[t]
\begin{center}
	\includegraphics[width=1\columnwidth]{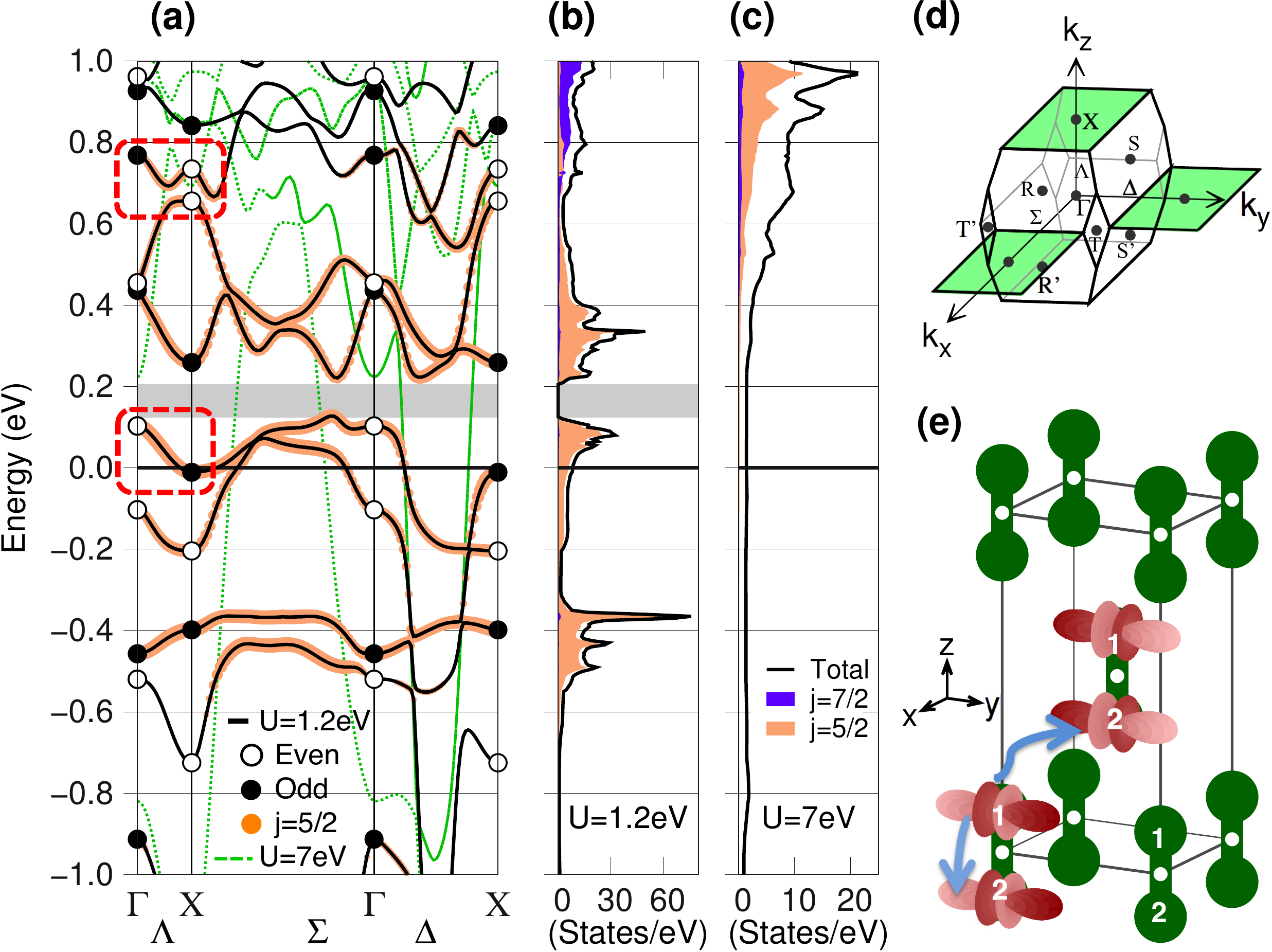}
\end{center}
\vspace{-3mm}
	\caption{(a) Two DFT+$U$ bands: $U$=1.2 eV (black solid lines) and 7 eV
(green line) where $5f$
electrons do and do not participate in the Fermi surface formation, respectively.   For the former,
the even/odd band parity and $j$=5/2 component are shown by the open/closed circles (black) and fat band representation (orange), respectively.  
The grey shaded region lying above the energy zero (chemical potential) represents a 81 meV bandgap
that separates a pair of topological $5f$ bands, indicated by by red-dotted boxes along the $\Lambda$
direction.  
%
%
Density of states for (b) $U$=1.2 eV and (c) 7 eV\@.  
(d) First BZ and eight TRIMs.  Additional X faces (green) of the neighboring BZs are shown to
demonstrate that all three principal $k$ lines ($\Lambda$, $\Sigma$, and $\Delta$) connect $\Gamma$
and X\@.
(e)  Wave function schematics of topological band at X point with odd parity.   
Thin/thick shading of lobes represents their positive/negative sign. 
In accord with the wave vector $k$=(0,0,2$\pi$), the sign array of body-centered orbitals is reversed from that of corner-centered.  
%
}
	\label{Fig:DFT}
\end{figure}

Many important questions remain open in understanding this 
  material, foremost of which is the origin of the odd-parity superconductivity. There is a consensus that ferromagnetic fluctuations are responsible
for the pairing in the related UGe$_2$, URhGe, and UCoGe compounds, \cite{Aoki:2019}, but uncertainty as to the
appropriate underlying model has led to debate over the nature of these ferromagnetic fluctuations
\cite{Aoki:2019}. Recently, this question
has been addressed in UTe$_2$ \cite{Shick:2019,Xu:2019,Ishizuka:2019} where density functional
theory plus Hubbard $U$ (DFT+$U$) and dynamical mean field theory (DMFT) calculations have developed
a family of band structures that depend upon $U$. The consequences of this family of band structures
on superconductivity have been explored, suggesting topological superconductivity
\cite{Ishizuka:2019}. In addition, effective Heisenberg theories have been developed, with the
insight that the strongest magnetic interaction, for all $U$ values considered, is a ferromagnetic interaction between the two 
nearest-neighbor 
U atoms on a ladder rung \cite{Xu:2019}, providing a potential mechanism for superconductivity. 

Here we revisit the DFT+$U$ calculations, finding good agreement with
previous results and newly
identifying a topological band that appears near the chemical potential for all values of $U$. This
topological band has its origin in the $5f$ electrons located predominantly on the rung sublattice
degrees of freedom. On the two U atoms of the rung, a band inversion between even- and odd-parity orbital combinations
provides the origin of the topological band. The appearance of the topological band together with
the rung ferromagnetic interaction discussed above indicates that the rung sublattice degrees of
freedom play a central role in the electronic description. Consequently,  we construct a
symmetry-based electronic model that explicitly includes these rung sublattice degrees of freedom
and the ferromagnetic interaction between them. This model yields magnetic field-temperature phase
diagrams that agree with experiment, allows a superconducting state
with Weyl nodes, and provides  an explanation for the observed surface chiral edge states \cite{Jiao:2020}, the observed polar Kerr effect \cite{Hayes:2020}, and observed low energy excitations in the superconducting state \cite{Metz:2019}.

\noindent {\it Topological Band}: A likely scenario for the electronic structure
of UTe$_2$ is a low temperature renormalized Fermi liquid ground state in which
U $5f$ electrons are participants. This is consistent with scanning tunneling
microscopy \cite{Jiao:2020} and the observed Fermi pocket about the $X$-point
seen in ARPES data \cite{Fujimori:2019,Miao:2020} (called $Z$ in
Ref.~\onlinecite{Miao:2020}).  This point of view has been adopted in recent
DMFT and DFT+$U$ calculations \cite{Shick:2019,Ishizuka:2019,Xu:2019}.  The
latter reveals that the band structure depends strongly on the choice of $U$,
suggesting that any theory of the superconducting state needs to be developed
for a range of band structures, emphasizing properties that are generic across
the relevant possibilities. Here, we have carried out DFT calculations of the
band structure of UTe$_2$ using the full-potential linearized augmented plane
wave method \cite{Mike_FLAPW_2009} and including a Coulomb $U$ to account for
interactions of the U $5f$ electrons. Our results agree with those found earlier \cite{Ishizuka:2019,Xu:2019}. A key new finding is that for all values of $U$ included here, we find a topological band at or near the chemical potential. 

As reported earlier, the band structure differs significantly along the three principal $k$ axes due
to the underlying quasi one-dimensional (1D) bands, whose features vividly
unfold in the no $5f$
electron limit or, equivalently, in the large $U$ limit as shown in
Fig.~\ref{Fig:DFT}(a).  The quasi-1D bands arise from the U $6d$ dimer
state which strongly disperses along the $k_x$ ($\Sigma$) direction,
and also the Te $5p$ linear chain state which disperses along the
  $k_y$ ($\Delta$) direction.  
For realistic values of $U$ (see the $U=1.2$eV band in
  Fig.~\ref{Fig:DFT}(a)), 
the $5f$ states are able to hybridize with these bands, leading to rather complicated
dispersions along $\Sigma$ and $\Delta$.    In
contrast, the $5f$ dispersion along $\Lambda$ is much simpler and we can make the following observations
for the $j=5/2$ sector:  
(i) Among the six  Kramers-degenerate bands,  two are  \textit{topologically} \textit{nontrivial} in the sense that the
band parity switches between $\Gamma$ and $X$, while the other four bands do not show such parity
change;      
(ii) these two are well separated in energy; 
(iii) the lower energy band, located near the chemical potential (energy zero), has even parity at
$\Gamma$ and odd at $X$; and  
(iv) of the four trivial bands, a set of odd- and even-parity bands
are occupied.
  
These features persistently exist regardless of $U$.   
The band structure and density of states (DOS) plots
[Fig.~\ref{Fig:DFT}(b)(c)], however, constrain the range of $U$
  that reproduce the experimental results such as
strong ARPES signals around $-$0.5 eV \cite{Fujimori:2019} or $-$0.7 eV \cite{Miao:2020}. 
In particular, for a range of moderate $U$ (1.1 eV $\le$ U $\le$ 2.0 eV), 
a band gap appears just above the lower nontrivial band; 
the Hilbert space below this gap (corresponding to the occupied levels of a $+2e$ doped system) is
characterized by $Z_2$ topological invariants $(\nu_0; \nu_1,\nu_2,\nu_3)$ \cite{FuKane:2007}, which
are found from the knowledge of band parity at eight time-reversal-invariant momenta (TRIMs).  Due
to a mirror symmetry duplication of $R$, $S$, and $T$ [c.f., Fig.~\ref{Fig:DFT}(d)], the index
$\nu_0$ is determined solely from the parity products at $\Gamma$ and $X$, $(-1)^{\nu_0} =
\delta_\Gamma \delta_X$ for our choice of origin. 
The $5f$ band with nontrivial parity switching leads to a strong topological state $\nu_0$=1.  
The other indices are all identical, $\nu_1$=$\nu_2$=$\nu_3$=1, determined from $(-1)^{\nu_1}=\delta_X\delta_R\delta_S\delta_T$.  
In a smaller $U$ range that includes $U$=0 eV, the non-doped system now has a
genuine insulating band gap where exactly the same $5f$ band provides $\nu_0$=1. 
The topological $5f$ band in focus is predominantly comprised
of $y(5y^2-3r^2)$ orbitals  on each of the two $U$ atoms forming
  a rung.
 At
the $\Gamma$ point, the wavefunction has opposite sign on these two
atoms, and hence has positive parity; as
sketched in Fig.~\ref{Fig:DFT}(e), however, at the $X$ point the
wavefunction has the same sign on the two rung atoms and is therefore odd
parity. 
 Common to both $k$ points is the U(1)-U(2) bond that connects different lattice points (different rungs).
For the $U$ used here (1.2 eV), this topological band gives rise to a Fermi surface that is centered on the $X$-point, in agreement with the Fermi pocket observed experimentally \cite{Fujimori:2019,Miao:2020}. 
More details of band structure analysis including the $U$ dependency are found in Supplemental Material.
 
\begin{figure}
\begin{center}
	\includegraphics[width=\columnwidth]{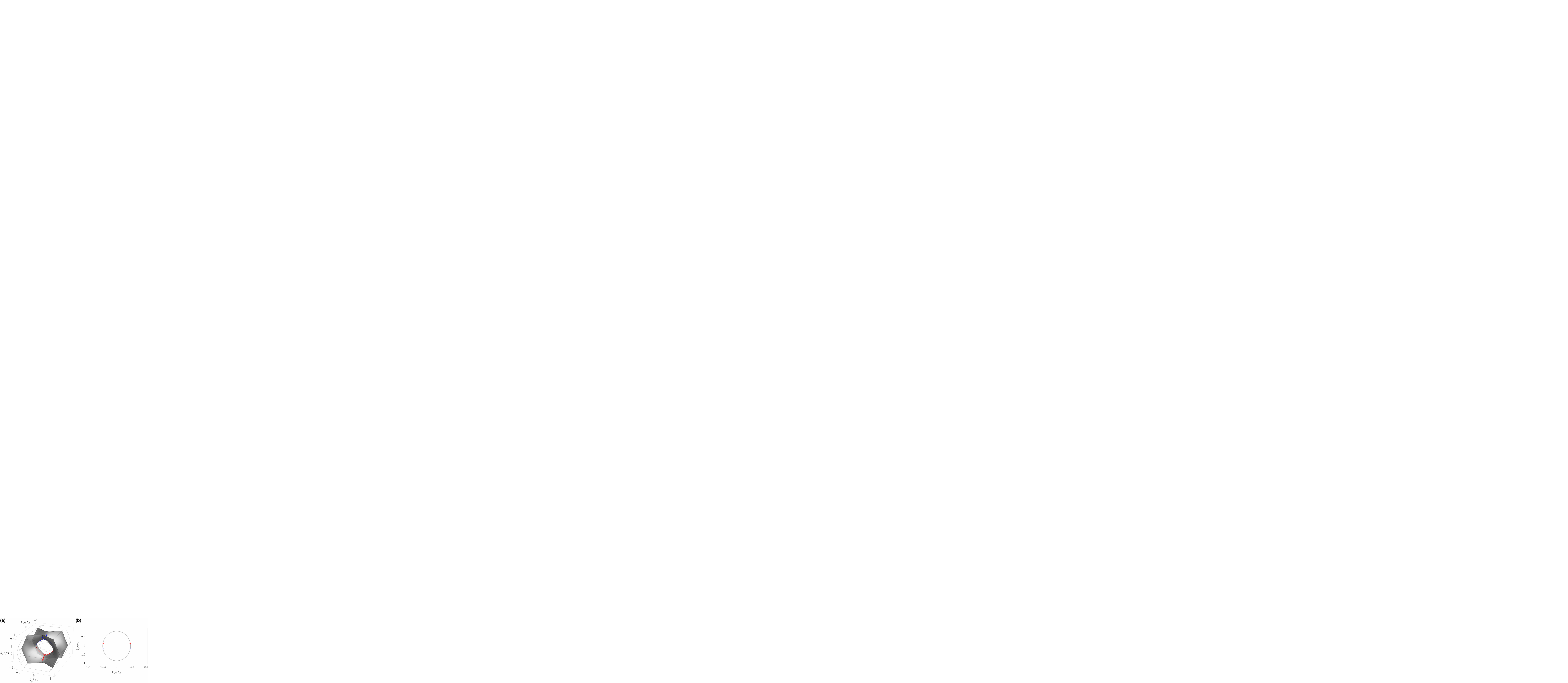}
\end{center}
	\caption{
		(a)  Evolution of Weyl nodes of $B_{3u}+iB_{2u}$ pairing states by decreasing the relative amplitude of the $B_{2u}$ gap pairing to the $B_{3u}$ gap pairing.
		The gap amplitudes are chosen to be $(\Delta_x,\Delta_y)=\Delta_0(\cos\theta,\sin\theta)$ where $\theta$ is a real parameter. The Fermi surface is obtained from the tight-binding model fitted to the two DFT+$U$ bands with $U$=1.2 eV. Red and blue lines indicate trajectories of the nodes with +1 and $-$1 Weyl charge as $\theta$ is varied. There are four nodes in a Brillouin zone and they sit on either  the $k_x$-$k_z$ or the $k_y$-$k_z$ plane. (b) Weyl points on a $k_x$-$k_z$ slice. Triangles(blue) and circles(red) indicate +1 and $-$1 Weyl charge. The circular line is a cut of the Fermi surface and centered at the $X$-point.
} \label{Fig:Weyl}
\end{figure}

\noindent {\it Minimal Hamiltonian:} The topological band and the ferromagnetic rung interaction
found by DFT~\cite{Xu:2019} indicates that the sublattice degree of
freedom due to the U atoms on a rung plays an important role in
the low-energy  physics. Surprisingly, this degree of freedom has not been explicitly considered
previously in understanding the superconducting state in UTe$_2$,
nor
in the related materials
UGe$_2$, URhGe, and UCoGe where a similar U sublattice structure
  appears  \cite{Aoki:2019}. Here we
consider the role of  this sublattice degree of freedom through the construction of a minimal model. In particular, the U atoms sit on
sites of $C_{2v}$ symmetry, for which only a single spinor symmetry representation exists. A minimal
model therefore includes a single spinor pair centered on each of the
 sublattices.   While these spinors share the same symmetry properties as usual spin-1/2 fermions  under
$C_{2v}$ symmetry, DFT reveals they are generally a linear combination
of $j=5/2$ states. 
This model only includes two bands. However, we note ARPES measurements have observed only one Fermi pocket associated with the U $5f$ electrons \cite{Miao:2020}, which are responsible for the superconducting state, suggesting this model is a reasonable description.
 The most general noninteracting Hamiltonian
  including all symmetry-allowed terms with sublattice and spin
  degrees of freedom is
\begin{eqnarray}
H_N= &&\epsilon_0(k)-\mu +f_{A_g}(k)\tau_x +f_z(k)\tau_y+f_y(k)\sigma_x\tau_z \nonumber \\
  && +f_x(k)\sigma_y\tau_z+f_{A_u}(k)\sigma_z\tau_z
\end{eqnarray}
where the functions $f_i(k)$ carry the symmetry properties given by the label $i$, in particular
$f_{A_g}(k)\sim$ constant, $f_z(k)\sim k_z$, $f_y(k)\sim k_y$, $f_x(k)\sim k_x$, and
$f_{A_u}(k) \sim k_xk_yk_z$. Here the Pauli matrices $\sigma_i$
describe the spin degrees of freedom and the Pauli matrices $\tau_i$
describe the rung degrees of freedom. While our analysis below does
not depend upon the detailed form of  the $f_i(k)$,  for
  the $y(5y^2-3r^2)$ orbitals discussed above we obtain the following tight-binding theory
\begin{eqnarray}
  \epsilon_0(k)&=&t_1\cos(k_x)+t_2\cos(k_y) \nonumber\\
  f_{A_g}(k)&=& m_0+t_3\cos(k_x/2)\cos(k_y/2)\cos(k_z/2) \nonumber \\
   f_z(k)&=&t_z\sin(k_z/2)\cos(k_x/2)\cos(k_y/2)\nonumber \\
   f_y(k)&=&t_y\sin(k_y) \nonumber \\
   f_x(k)&=&t_x\sin(k_x)\nonumber\\
   f_{A_u}(k)&=& t_u \sin(k_x/2)\sin(k_y/2)\sin(k_z/2) \ .
\end{eqnarray}
Note that to  replicate the nontrivial parity switching predicted
above, the magnitude of the \textit{inter-rung} U(1)-U(2) hopping
$t_3$ [see Fig.~\ref{Fig:DFT}(e)] needs to exceed the \textit{intra-rung} hopping $m_0$. Fitting to the DFT band near the Fermi surface gives 
$(\mu,t_1,t_2,m_0,t_3,t_z,t_x,t_y,t_u)$ = (0.129, $-$0.0892, 0.0678,
$-$0.062, 0.0742, $-$0.0742, 0.006, 0.008, 0.01). This fit yields the Fermi surface shown in Fig~\ref{Fig:Weyl}. 

\noindent {\it Magnetic interactions:}  DFT calculations have found that the dominant magnetic
interaction is a ferromagnetic interaction between the rung sublattice U atoms \cite{Xu:2019}.  Note
that this local ferromagnetic interaction does not imply a global ferromagnetic state, but only that
these two U atoms have the same spin-orientation. Indeed, DFT finds ferromagnetic and
anti-ferromagnetic ground states consistent with this local
configuration \cite{Xu:2019},  which  may account for  the two magnetically-ordered states observed experimentally \cite{Thomas:2020}.  This interaction is given by 
\begin{equation}
H_{int}=-\sum_{i}(J_x S^x_{i,1}{S}^x_{i,2}+J_y S^y_{i,1}{S}^y_{i,2}+J_z S^z_{i,1}S^z_{i,2})
\end{equation}
where $1,2$ labels the two U atoms on the rung, and $i$ labels a lattice point; the ferromagnetic interactions $J_\mu>0$ are in general unequal due to the
orthorhombic structure. Treating this as an effective coupling for superconductivity, we find this
gives rise to three possible pairing states as  listed in Table~\ref{Tab1}.
Due to the inter-sublattice nature of the magnetic interactions, the
gap functions are necessarily proportional to a non-trivial $\tau_y$ sublattice  operator and take the form $\Delta_i
\tau_y \sigma_i$ which describes a local, inter-sublattice, spin-triplet pairing function. 
While the interactions reveal the role of magnetic anisotropy on pairing interaction, we will now set $J_x=J_y=J_z$ to examine the effect of $H_N$ on these pairing states.

\begin{table}
\caption{Pairing gap functions due to ferromagnetic interactions between rung sublattice degrees of
freedom. The first column gives the local gap function and the last column gives the corresponding
$\vec{d}(k)$ in the band basis when the spin-orbit coupling terms are vanishing ($f_x=f_y=f_{A_u}=0$).
 \label{Tab1}}
\begin{ruledtabular}
\begin{tabular}{cccc}
  Gap & Irrep& Interaction& Momentum dependence\\
  \hline
   &  \\ [-7pt]  
   $\Delta_z\tau_y\sigma_z$& $A_{u}$& $\phantom{-}J_x+J_y-J_z$ &
     $\displaystyle \frac{f_z(k)}{\sqrt{f_{A_g}^2(k)+f_z^2(k)}}\,\hat{z}$ \\
  $\Delta_x\tau_y\sigma_x$& $B_{2u}$& $-  J_x+J_y+J_z$ &
     $\displaystyle \frac{f_z(k)}{\sqrt{f_{A_g}^2(k)+f_z^2(k)}}\,\hat{x}$  \\
  $\Delta_y\tau_y\sigma_y$& $B_{3u}$& $\phantom{-}J_x-J_y+J_z$ & 
     $\displaystyle \frac{f_z(k)}{\sqrt{f_{A_g}^2(k)+f_z^2(k)}}\, \hat{y}$\\
  \end{tabular}
\end{ruledtabular}
\end{table}

\noindent {\it Role of $H_N$:} Naively, the stable pairing state is determined by the largest
interaction parameter listed in Table~\ref{Tab1}. However,  due
  to the spin-sublattice coupling in our model,  $H_N$ also influences the
relative stability of the pairing states.  The effect of the
  distinct terms in $H_N$ on the transition temperature $T_{c,i}$ of
  the state $\Delta_i \tau_y \sigma_i$ 
  can be quantified without fully specifying the functions $f_j(k)$
  using the concept of superconducting fitness \cite{Ramires:2016,Ramires:2018}: specifically, if the matrix $\sigma_i\tau_j$ anticommutes with the
gap function
$\Delta_i\tau_y \sigma_i$, then the corresponding term in $H_N$ will
enhance $T_{c,i}$; conversely, $T_{c,i}$ is suppressed by this term if
it commutes with the pairing potential \cite{Ramires:2018}. This yields the result that the $f_{A_g}$ term suppresses all the $T_{c,i}$ and the $f_z$ term enhances all the $T_{c,i}$. 
Consequently, if $J_x=J_y=J_z$,  the spin-orbit coupling  terms will dictate which $T_{c,i}$ is
highest. In particular, the largest $T_{c,i}$ is given by the smallest
of $\langle f_{A_u}^2\rangle$ ($A_{u}$ stable), $\langle
f_{x}^2\rangle$ ($B_{3u}$ stable), or  $\langle f_{y}^2\rangle$
($B_{2u}$ stable), where $\langle \ldots \rangle$ represents an average
over the Fermi surface.  The terms in the tight-binding expression will be altered by pressure, providing a potential explanation for the appearance of different superconducting states.

$H_N$ also dictates the form of the pseudospin triplet $\vec{d}$-vector on the Fermi surface. We do
not give the details here but point out that generically, all three pseudospin components
$\hat{x},\hat{y},\hat{z}$ appear for each gap function. As we argue below, there is one limit that can be motivated by experimental results. In particular, when the momentum-dependent
spin-orbit coupling terms are small, that is $f_x^2$, $f_y^2$, $f_{Au}^2 << f_z^2$  then
the orientation of the
spin-triplet $\vec{d}$ vectors  is set by the spin part of the
  gap function  in Table~\ref{Tab1}.   In this case Table~\ref{Tab1} provides an
approximately correct description of $\vec{d}$ (except near $k_z=0,2\pi$); in the following this is
called the weak spin-orbit coupling limit.

\noindent{\it Relationship to experiment:} At ambient pressure, two superconducting transitions in zero field and a polar Kerr effect that can be trained by a $c$-axis magnetic field has been observed \cite{Hayes:2020}.  The latter result implies a $B_{3u}+iB_{2u}$ or a $A_{u}+iB_{1u}$ pairing below the second transition \cite{Hayes:2020}. In the context of our theory, the only possibility is the  $B_{3u}+iB_{2u}$ state.  Such a broken-time reversal symmetry state can be stabilized by ferromagnetic fluctuations \cite{Hayes:2020,Yarzhemsky:2020,Nevidomskyy:2020}. Assuming an isotropic rung
exchange, this situation arises in our model by requiring that $\langle
f_{x}^2\rangle<\langle f_{y}^2\rangle < \langle f_{A_u}^2\rangle$. 

 We note that additional consistency with experiment arises if we assume we are in the weak spin-orbit coupling limit. In particular, this limit naturally explains why  thermal conductivity
exhibits nodal behavior that is similar along both the $\hat{a}$ and $\hat{b}$ directions
\cite{Metz:2019}: When $f_x=f_y=f_{A_u}=0$, all the gap functions have
accidental line nodes when $k_z=0$, yielding nodal thermal
conductivity behavior along both $\hat{a}$ and $\hat{b}$.  These
accidental line nodes will be lifted when the spin-orbit coupling
terms are non-zero,  but if they are small we expect  a local gap minimum near $k_z=0,2\pi$
which can mimic nodes in thermal conductivity. 

\begin{figure}
\begin{center}
	\includegraphics[width=\columnwidth]{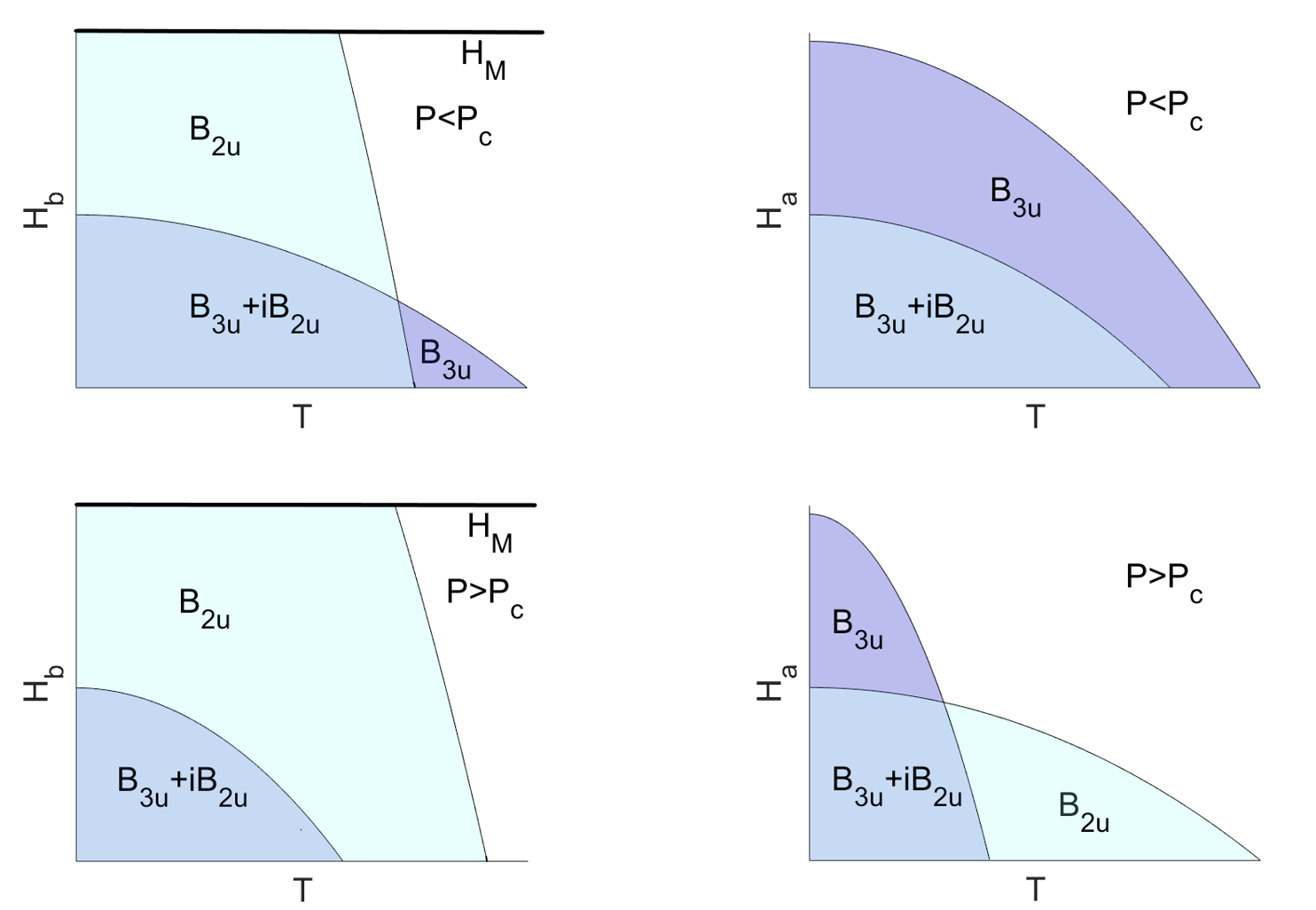}
\end{center}
	\caption{Qualitative temperature-field phase diagrams for fields along the $\hat{a}$ and $\hat{b}$ directions. The top two phase diagrams correspond to $P<0.2$ GPa and the bottom two to $P>0.2$ GPa.  $H_M$ corresponds to an observed metamagnetic transition. 
}\label{Fig:Phase}
\end{figure}

 In addition, the weak spin-orbit coupling limit is consistent with the field dependence of the phase diagram for fields along the $\hat{a}$ and $\hat{b}$ as a function of pressure \cite{Braithwaite:2019,Thomas:2020,Aoki:2020,Lin:2020}. In particular, in this limit the $B_{3u}$ gap is primarily along $\hat{b}$ and the $B_{2u}$ gap is primarily along $\hat{a}$. This implies that the $B_{3u}$ ($B_{2u}$) gap will experience paramagnetic limiting for a field along $\hat{b}$ ($\hat{a}$) and not for the fields along $\hat{a}$ ($\hat{b}$). It has been observed that the two superconducting transitions cross at a critical pressure $P_c\approx 0.2$ GPa \cite{Thomas:2020}. In our model such a crossing should then be correlated with a switch in the upper critical field behavior for the field along the $\hat{a}$ and $\hat{b}$ directions, as shown in Fig.~\ref{Fig:Phase}. This is indeed what is observed \cite{Ran:2019,Aoki:2020,Lin:2020}.

\noindent {\it Weyl Nodes:}  Here we examine more carefully the nodal
structure of a  $B_{3u}+iB_{2u}$ pairing state. Using the tight-binding theory given above, we find
that Weyl nodes generically exist. These nodes are topologically protected but  do not sit at
positions of high symmetry. The position of these nodes are determined by the relative amplitudes of
the $B_{2u}$ and the $B_{3u}$ order parameters. The evolution of these nodes is shown in
Fig.~\ref{Fig:Weyl}. We have also computed the Weyl charge of these nodes. Generically, there exists
four Weyl nodes, two of charge +1 and two of charge $-$1. These Weyl nodes imply the existence of
surface Fermi arc states which provide an explanation for the chiral edges states seen with scanning
tunneling microscopy \cite{Jiao:2020}.

\noindent {\it Polar Kerr Effect:}  Our multiband theory for the
superconductivity in UTe$_2$ generically gives rise to an imaginary anomalous Hall conductivity,
which is expected to be proportional to the polar Kerr signal. By a sum rule~\cite{Lange:1999}
we have that the integrated
imaginary anomalous Hall conductivity is given by $\int^\infty_{-\infty}
\omega\sigma_{H}(\omega)d\omega = -i\pi e^2\langle [\partial_{k_x}
H_N, \partial_{k_y}H_N]\rangle$. The full expansion of the commutator
is very complicated and will be analyzed elsewhere, but we note that
the contribution
$(\partial_{k_x}f_y\partial_{k_y}f_x-\partial_{k_x}f_x\partial_{k_y}f_y)\sigma_z\tau_0$
is directly proportional to the so-called time-reversal-odd bilinear
of the $B_{3u}+iB_{2u}$ pairing state \cite{Brydon:2018,Brydon:2019}. This implies that 
expectation value of the commutator is nonzero, ensuring the existence
of the anomalous Hall conductivity and hence the polar Kerr
signal. The presence of two bands due to the sublattice degree
  of freedom is
critical to this argument; in a single-band model, the commutator is
vanishing,  and a polar Kerr effect does not appear in the clean limit~\cite{Taylor:2012}.

From DFT+$U$ calculations, we have identified a topological band near the chemical potential in UTe$_2$ that stems from U 5$f$ electrons. This result, together with the importance of rung ferromagnetic interactions, suggests that U atom rung
degrees of freedom play an important role in superconducting UTe$_2$. We have  developed a model that includes these degrees of freedom and captures the topological bands. In addition, we show that including the ferromagnetic rung interactions allows a $B_{3u}+iB_{2u}$ pairing state, accounting for Polar Kerr measurements and yielding Weyl points, providing
a promising model with which to understand UTe$_2$ in more detail. Similar U sublattice degrees of freedom exist in UGe$_2$, URhGe, and UCoGe, suggesting a unifying motif for this class of materials. 

We acknowledge useful discussions with Nick Butch, Yun Suk Eo, Ian Hayes,  Aharon Kapitulnik, Johnpierre Paglione, Srinivas Raghu, and Di Wei. This work was supported by
the U.S. Department of Energy, Office of Basic Energy Sciences,
Division of Materials Sciences and Engineering under Award
DE-SC0017632.  P.M.R.B. was supported by the Marsden Fund Council
  from Government funding, managed by Royal Society Te Ap\={a}rangi.

\bibliography{UTe2paper}

\begin{thebibliography}{39}%
\makeatletter
\providecommand \@ifxundefined [1]{%
 \@ifx{#1\undefined}
}%
\providecommand \@ifnum [1]{%
 \ifnum #1\expandafter \@firstoftwo
 \else \expandafter \@secondoftwo
 \fi
}%
\providecommand \@ifx [1]{%
 \ifx #1\expandafter \@firstoftwo
 \else \expandafter \@secondoftwo
 \fi
}%
\providecommand \natexlab [1]{#1}%
\providecommand \enquote  [1]{``#1''}%
\providecommand \bibnamefont  [1]{#1}%
\providecommand \bibfnamefont [1]{#1}%
\providecommand \citenamefont [1]{#1}%
\providecommand \href@noop [0]{\@secondoftwo}%
\providecommand \href [0]{\begingroup \@sanitize@url \@href}%
\providecommand \@href[1]{\@@startlink{#1}\@@href}%
\providecommand \@@href[1]{\endgroup#1\@@endlink}%
\providecommand \@sanitize@url [0]{\catcode `\\12\catcode `\$12\catcode
  `\&12\catcode `\#12\catcode `\^12\catcode `\_12\catcode `\%12\relax}%
\providecommand \@@startlink[1]{}%
\providecommand \@@endlink[0]{}%
\providecommand \url  [0]{\begingroup\@sanitize@url \@url }%
\providecommand \@url [1]{\endgroup\@href {#1}{\urlprefix }}%
\providecommand \urlprefix  [0]{URL }%
\providecommand \Eprint [0]{\href }%
\providecommand \doibase [0]{http://dx.doi.org/}%
\providecommand \selectlanguage [0]{\@gobble}%
\providecommand \bibinfo  [0]{\@secondoftwo}%
\providecommand \bibfield  [0]{\@secondoftwo}%
\providecommand \translation [1]{[#1]}%
\providecommand \BibitemOpen [0]{}%
\providecommand \bibitemStop [0]{}%
\providecommand \bibitemNoStop [0]{.\EOS\space}%
\providecommand \EOS [0]{\spacefactor3000\relax}%
\providecommand \BibitemShut  [1]{\csname bibitem#1\endcsname}%
\let\auto@bib@innerbib\@empty
\bibitem [{\citenamefont {Ran}\ \emph {et~al.}(2019{\natexlab{a}})\citenamefont
  {Ran}, \citenamefont {Eckberg}, \citenamefont {Ding}, \citenamefont
  {Furukawa}, \citenamefont {Metz}, \citenamefont {Saha}, \citenamefont {Liu},
  \citenamefont {Zic}, \citenamefont {Kim}, \citenamefont {Paglione},\ and\
  \citenamefont {Butch}}]{Ran:2019}%
  \BibitemOpen
  \bibfield  {author} {\bibinfo {author} {\bibfnamefont {S.}~\bibnamefont
  {Ran}}, \bibinfo {author} {\bibfnamefont {C.}~\bibnamefont {Eckberg}},
  \bibinfo {author} {\bibfnamefont {Q.-P.}\ \bibnamefont {Ding}}, \bibinfo
  {author} {\bibfnamefont {Y.}~\bibnamefont {Furukawa}}, \bibinfo {author}
  {\bibfnamefont {T.}~\bibnamefont {Metz}}, \bibinfo {author} {\bibfnamefont
  {S.~R.}\ \bibnamefont {Saha}}, \bibinfo {author} {\bibfnamefont {I.-L.}\
  \bibnamefont {Liu}}, \bibinfo {author} {\bibfnamefont {M.}~\bibnamefont
  {Zic}}, \bibinfo {author} {\bibfnamefont {H.}~\bibnamefont {Kim}}, \bibinfo
  {author} {\bibfnamefont {J.}~\bibnamefont {Paglione}}, \ and\ \bibinfo
  {author} {\bibfnamefont {N.~P.}\ \bibnamefont {Butch}},\ }\href {\doibase
  10.1126/science.aav8645} {\bibfield  {journal} {\bibinfo  {journal}
  {Science}\ }\textbf {\bibinfo {volume} {365}},\ \bibinfo {pages} {684}
  (\bibinfo {year} {2019}{\natexlab{a}})}\BibitemShut {NoStop}%
\bibitem [{\citenamefont {Aoki}\ \emph
  {et~al.}(2019{\natexlab{a}})\citenamefont {Aoki}, \citenamefont {Nakamura},
  \citenamefont {Honda}, \citenamefont {Li}, \citenamefont {Homma},
  \citenamefont {Shimizu}, \citenamefont {Sato}, \citenamefont {Knebel},
  \citenamefont {Brison}, \citenamefont {Pourret}, \citenamefont {Braithwaite},
  \citenamefont {Lapertot}, \citenamefont {Niu}, \citenamefont {Vališka},
  \citenamefont {Harima},\ and\ \citenamefont {Flouquet}}]{Aoki:2019-1}%
  \BibitemOpen
  \bibfield  {author} {\bibinfo {author} {\bibfnamefont {D.}~\bibnamefont
  {Aoki}}, \bibinfo {author} {\bibfnamefont {A.}~\bibnamefont {Nakamura}},
  \bibinfo {author} {\bibfnamefont {F.}~\bibnamefont {Honda}}, \bibinfo
  {author} {\bibfnamefont {D.}~\bibnamefont {Li}}, \bibinfo {author}
  {\bibfnamefont {Y.}~\bibnamefont {Homma}}, \bibinfo {author} {\bibfnamefont
  {Y.}~\bibnamefont {Shimizu}}, \bibinfo {author} {\bibfnamefont {Y.~J.}\
  \bibnamefont {Sato}}, \bibinfo {author} {\bibfnamefont {G.}~\bibnamefont
  {Knebel}}, \bibinfo {author} {\bibfnamefont {J.-P.}\ \bibnamefont {Brison}},
  \bibinfo {author} {\bibfnamefont {A.}~\bibnamefont {Pourret}}, \bibinfo
  {author} {\bibfnamefont {D.}~\bibnamefont {Braithwaite}}, \bibinfo {author}
  {\bibfnamefont {G.}~\bibnamefont {Lapertot}}, \bibinfo {author}
  {\bibfnamefont {Q.}~\bibnamefont {Niu}}, \bibinfo {author} {\bibfnamefont
  {M.}~\bibnamefont {Vališka}}, \bibinfo {author} {\bibfnamefont
  {H.}~\bibnamefont {Harima}}, \ and\ \bibinfo {author} {\bibfnamefont
  {J.}~\bibnamefont {Flouquet}},\ }\href {\doibase 10.7566/JPSJ.88.043702}
  {\bibfield  {journal} {\bibinfo  {journal} {Journal of the Physical Society
  of Japan}\ }\textbf {\bibinfo {volume} {88}},\ \bibinfo {pages} {043702}
  (\bibinfo {year} {2019}{\natexlab{a}})},\ \Eprint
  {http://arxiv.org/abs/https://doi.org/10.7566/JPSJ.88.043702}
  {https://doi.org/10.7566/JPSJ.88.043702} \BibitemShut {NoStop}%
\bibitem [{\citenamefont {Knafo}\ \emph {et~al.}(2019)\citenamefont {Knafo},
  \citenamefont {Vališka}, \citenamefont {Braithwaite}, \citenamefont
  {Lapertot}, \citenamefont {Knebel}, \citenamefont {Pourret}, \citenamefont
  {Brison}, \citenamefont {Flouquet},\ and\ \citenamefont {Aoki}}]{Knafo:2019}%
  \BibitemOpen
  \bibfield  {author} {\bibinfo {author} {\bibfnamefont {W.}~\bibnamefont
  {Knafo}}, \bibinfo {author} {\bibfnamefont {M.}~\bibnamefont {Vališka}},
  \bibinfo {author} {\bibfnamefont {D.}~\bibnamefont {Braithwaite}}, \bibinfo
  {author} {\bibfnamefont {G.}~\bibnamefont {Lapertot}}, \bibinfo {author}
  {\bibfnamefont {G.}~\bibnamefont {Knebel}}, \bibinfo {author} {\bibfnamefont
  {A.}~\bibnamefont {Pourret}}, \bibinfo {author} {\bibfnamefont {J.-P.}\
  \bibnamefont {Brison}}, \bibinfo {author} {\bibfnamefont {J.}~\bibnamefont
  {Flouquet}}, \ and\ \bibinfo {author} {\bibfnamefont {D.}~\bibnamefont
  {Aoki}},\ }\href {\doibase 10.7566/JPSJ.88.063705} {\bibfield  {journal}
  {\bibinfo  {journal} {Journal of the Physical Society of Japan}\ }\textbf
  {\bibinfo {volume} {88}},\ \bibinfo {pages} {063705} (\bibinfo {year}
  {2019})},\ \Eprint
  {http://arxiv.org/abs/https://doi.org/10.7566/JPSJ.88.063705}
  {https://doi.org/10.7566/JPSJ.88.063705} \BibitemShut {NoStop}%
\bibitem [{\citenamefont {Miyake}\ \emph {et~al.}(2019)\citenamefont {Miyake},
  \citenamefont {Shimizu}, \citenamefont {Sato}, \citenamefont {Li},
  \citenamefont {Nakamura}, \citenamefont {Homma}, \citenamefont {Honda},
  \citenamefont {Flouquet}, \citenamefont {Tokunaga},\ and\ \citenamefont
  {Aoki}}]{Miyake:2019}%
  \BibitemOpen
  \bibfield  {author} {\bibinfo {author} {\bibfnamefont {A.}~\bibnamefont
  {Miyake}}, \bibinfo {author} {\bibfnamefont {Y.}~\bibnamefont {Shimizu}},
  \bibinfo {author} {\bibfnamefont {Y.~J.}\ \bibnamefont {Sato}}, \bibinfo
  {author} {\bibfnamefont {D.}~\bibnamefont {Li}}, \bibinfo {author}
  {\bibfnamefont {A.}~\bibnamefont {Nakamura}}, \bibinfo {author}
  {\bibfnamefont {Y.}~\bibnamefont {Homma}}, \bibinfo {author} {\bibfnamefont
  {F.}~\bibnamefont {Honda}}, \bibinfo {author} {\bibfnamefont
  {J.}~\bibnamefont {Flouquet}}, \bibinfo {author} {\bibfnamefont
  {M.}~\bibnamefont {Tokunaga}}, \ and\ \bibinfo {author} {\bibfnamefont
  {D.}~\bibnamefont {Aoki}},\ }\href {\doibase 10.7566/JPSJ.88.063706}
  {\bibfield  {journal} {\bibinfo  {journal} {Journal of the Physical Society
  of Japan}\ }\textbf {\bibinfo {volume} {88}},\ \bibinfo {pages} {063706}
  (\bibinfo {year} {2019})},\ \Eprint
  {http://arxiv.org/abs/https://doi.org/10.7566/JPSJ.88.063706}
  {https://doi.org/10.7566/JPSJ.88.063706} \BibitemShut {NoStop}%
\bibitem [{\citenamefont {Imajo}\ \emph {et~al.}(2019)\citenamefont {Imajo},
  \citenamefont {Kohama}, \citenamefont {Miyake}, \citenamefont {Dong},
  \citenamefont {Tokunaga}, \citenamefont {Flouquet}, \citenamefont {Kindo},\
  and\ \citenamefont {Aoki}}]{Imajo:2019}%
  \BibitemOpen
  \bibfield  {author} {\bibinfo {author} {\bibfnamefont {S.}~\bibnamefont
  {Imajo}}, \bibinfo {author} {\bibfnamefont {Y.}~\bibnamefont {Kohama}},
  \bibinfo {author} {\bibfnamefont {A.}~\bibnamefont {Miyake}}, \bibinfo
  {author} {\bibfnamefont {C.}~\bibnamefont {Dong}}, \bibinfo {author}
  {\bibfnamefont {M.}~\bibnamefont {Tokunaga}}, \bibinfo {author}
  {\bibfnamefont {J.}~\bibnamefont {Flouquet}}, \bibinfo {author}
  {\bibfnamefont {K.}~\bibnamefont {Kindo}}, \ and\ \bibinfo {author}
  {\bibfnamefont {D.}~\bibnamefont {Aoki}},\ }\href {\doibase
  10.7566/JPSJ.88.083705} {\bibfield  {journal} {\bibinfo  {journal} {Journal
  of the Physical Society of Japan}\ }\textbf {\bibinfo {volume} {88}},\
  \bibinfo {pages} {083705} (\bibinfo {year} {2019})},\ \Eprint
  {http://arxiv.org/abs/https://doi.org/10.7566/JPSJ.88.083705}
  {https://doi.org/10.7566/JPSJ.88.083705} \BibitemShut {NoStop}%
\bibitem [{\citenamefont {Mineev}(2020)}]{Mineev:2020}%
  \BibitemOpen
  \bibfield  {author} {\bibinfo {author} {\bibfnamefont {V.~P.}\ \bibnamefont
  {Mineev}},\ }\href {\doibase 10.1134/s0021364020120036} {\bibfield  {journal}
  {\bibinfo  {journal} {JETP Letters}\ } (\bibinfo {year} {2020}),\
  10.1134/s0021364020120036}\BibitemShut {NoStop}%
\bibitem [{\citenamefont {Ran}\ \emph {et~al.}(2020)\citenamefont {Ran},
  \citenamefont {Kim}, \citenamefont {Liu}, \citenamefont {Saha}, \citenamefont
  {Hayes}, \citenamefont {Metz}, \citenamefont {Eo}, \citenamefont {Paglione},\
  and\ \citenamefont {Butch}}]{Ran:2020}%
  \BibitemOpen
  \bibfield  {author} {\bibinfo {author} {\bibfnamefont {S.}~\bibnamefont
  {Ran}}, \bibinfo {author} {\bibfnamefont {H.}~\bibnamefont {Kim}}, \bibinfo
  {author} {\bibfnamefont {I.-L.}\ \bibnamefont {Liu}}, \bibinfo {author}
  {\bibfnamefont {S.~R.}\ \bibnamefont {Saha}}, \bibinfo {author}
  {\bibfnamefont {I.}~\bibnamefont {Hayes}}, \bibinfo {author} {\bibfnamefont
  {T.}~\bibnamefont {Metz}}, \bibinfo {author} {\bibfnamefont {Y.~S.}\
  \bibnamefont {Eo}}, \bibinfo {author} {\bibfnamefont {J.}~\bibnamefont
  {Paglione}}, \ and\ \bibinfo {author} {\bibfnamefont {N.~P.}\ \bibnamefont
  {Butch}},\ }\href {\doibase 10.1103/PhysRevB.101.140503} {\bibfield
  {journal} {\bibinfo  {journal} {Phys. Rev. B}\ }\textbf {\bibinfo {volume}
  {101}},\ \bibinfo {pages} {140503} (\bibinfo {year} {2020})}\BibitemShut
  {NoStop}%
\bibitem [{\citenamefont {Knebel}\ \emph {et~al.}(2020)\citenamefont {Knebel},
  \citenamefont {Kimata}, \citenamefont {Vališka}, \citenamefont {Honda},
  \citenamefont {Li}, \citenamefont {Braithwaite}, \citenamefont {Lapertot},
  \citenamefont {Knafo}, \citenamefont {Pourret}, \citenamefont {Sato},
  \citenamefont {Shimizu}, \citenamefont {Kihara}, \citenamefont {Brison},
  \citenamefont {Flouquet},\ and\ \citenamefont {Aoki}}]{Knebel:2020}%
  \BibitemOpen
  \bibfield  {author} {\bibinfo {author} {\bibfnamefont {G.}~\bibnamefont
  {Knebel}}, \bibinfo {author} {\bibfnamefont {M.}~\bibnamefont {Kimata}},
  \bibinfo {author} {\bibfnamefont {M.}~\bibnamefont {Vališka}}, \bibinfo
  {author} {\bibfnamefont {F.}~\bibnamefont {Honda}}, \bibinfo {author}
  {\bibfnamefont {D.}~\bibnamefont {Li}}, \bibinfo {author} {\bibfnamefont
  {D.}~\bibnamefont {Braithwaite}}, \bibinfo {author} {\bibfnamefont
  {G.}~\bibnamefont {Lapertot}}, \bibinfo {author} {\bibfnamefont
  {W.}~\bibnamefont {Knafo}}, \bibinfo {author} {\bibfnamefont
  {A.}~\bibnamefont {Pourret}}, \bibinfo {author} {\bibfnamefont {Y.~J.}\
  \bibnamefont {Sato}}, \bibinfo {author} {\bibfnamefont {Y.}~\bibnamefont
  {Shimizu}}, \bibinfo {author} {\bibfnamefont {T.}~\bibnamefont {Kihara}},
  \bibinfo {author} {\bibfnamefont {J.-P.}\ \bibnamefont {Brison}}, \bibinfo
  {author} {\bibfnamefont {J.}~\bibnamefont {Flouquet}}, \ and\ \bibinfo
  {author} {\bibfnamefont {D.}~\bibnamefont {Aoki}},\ }\href {\doibase
  10.7566/JPSJ.89.053707} {\bibfield  {journal} {\bibinfo  {journal} {Journal
  of the Physical Society of Japan}\ }\textbf {\bibinfo {volume} {89}},\
  \bibinfo {pages} {053707} (\bibinfo {year} {2020})},\ \Eprint
  {http://arxiv.org/abs/https://doi.org/10.7566/JPSJ.89.053707}
  {https://doi.org/10.7566/JPSJ.89.053707} \BibitemShut {NoStop}%
\bibitem [{\citenamefont {Niu}\ \emph {et~al.}(2020)\citenamefont {Niu},
  \citenamefont {Knebel}, \citenamefont {Braithwaite}, \citenamefont {Aoki},
  \citenamefont {Lapertot}, \citenamefont {Seyfarth}, \citenamefont {Brison},
  \citenamefont {Flouquet},\ and\ \citenamefont {Pourret}}]{Niu:2020}%
  \BibitemOpen
  \bibfield  {author} {\bibinfo {author} {\bibfnamefont {Q.}~\bibnamefont
  {Niu}}, \bibinfo {author} {\bibfnamefont {G.}~\bibnamefont {Knebel}},
  \bibinfo {author} {\bibfnamefont {D.}~\bibnamefont {Braithwaite}}, \bibinfo
  {author} {\bibfnamefont {D.}~\bibnamefont {Aoki}}, \bibinfo {author}
  {\bibfnamefont {G.}~\bibnamefont {Lapertot}}, \bibinfo {author}
  {\bibfnamefont {G.}~\bibnamefont {Seyfarth}}, \bibinfo {author}
  {\bibfnamefont {J.-P.}\ \bibnamefont {Brison}}, \bibinfo {author}
  {\bibfnamefont {J.}~\bibnamefont {Flouquet}}, \ and\ \bibinfo {author}
  {\bibfnamefont {A.}~\bibnamefont {Pourret}},\ }\href {\doibase
  10.1103/PhysRevLett.124.086601} {\bibfield  {journal} {\bibinfo  {journal}
  {Phys. Rev. Lett.}\ }\textbf {\bibinfo {volume} {124}},\ \bibinfo {pages}
  {086601} (\bibinfo {year} {2020})}\BibitemShut {NoStop}%
\bibitem [{\citenamefont {Ran}\ \emph {et~al.}(2019{\natexlab{b}})\citenamefont
  {Ran}, \citenamefont {Liu}, \citenamefont {Eo}, \citenamefont {Campbell},
  \citenamefont {Neves}, \citenamefont {Fuhrman}, \citenamefont {Saha},
  \citenamefont {Eckberg}, \citenamefont {Kim}, \citenamefont {Graf},
  \citenamefont {Balakirev}, \citenamefont {Singleton}, \citenamefont
  {Paglione},\ and\ \citenamefont {Butch}}]{Ran:2019-2}%
  \BibitemOpen
  \bibfield  {author} {\bibinfo {author} {\bibfnamefont {S.}~\bibnamefont
  {Ran}}, \bibinfo {author} {\bibfnamefont {I.-L.}\ \bibnamefont {Liu}},
  \bibinfo {author} {\bibfnamefont {Y.~S.}\ \bibnamefont {Eo}}, \bibinfo
  {author} {\bibfnamefont {D.~J.}\ \bibnamefont {Campbell}}, \bibinfo {author}
  {\bibfnamefont {P.~M.}\ \bibnamefont {Neves}}, \bibinfo {author}
  {\bibfnamefont {W.~T.}\ \bibnamefont {Fuhrman}}, \bibinfo {author}
  {\bibfnamefont {S.~R.}\ \bibnamefont {Saha}}, \bibinfo {author}
  {\bibfnamefont {C.}~\bibnamefont {Eckberg}}, \bibinfo {author} {\bibfnamefont
  {H.}~\bibnamefont {Kim}}, \bibinfo {author} {\bibfnamefont {D.}~\bibnamefont
  {Graf}}, \bibinfo {author} {\bibfnamefont {F.}~\bibnamefont {Balakirev}},
  \bibinfo {author} {\bibfnamefont {J.}~\bibnamefont {Singleton}}, \bibinfo
  {author} {\bibfnamefont {J.}~\bibnamefont {Paglione}}, \ and\ \bibinfo
  {author} {\bibfnamefont {N.~P.}\ \bibnamefont {Butch}},\ }\href {\doibase
  {10.1038/s41567-019-0670-x}} {\bibfield  {journal} {\bibinfo  {journal}
  {{Nature Physics}}\ }\textbf {\bibinfo {volume} {{15}}},\ \bibinfo {pages}
  {{1250}} (\bibinfo {year} {{2019}}{\natexlab{b}})}\BibitemShut {NoStop}%
\bibitem [{\citenamefont {Tokunaga}\ \emph {et~al.}(2019)\citenamefont
  {Tokunaga}, \citenamefont {Sakai}, \citenamefont {Kambe}, \citenamefont
  {Hattori}, \citenamefont {Higa}, \citenamefont {Nakamine}, \citenamefont
  {Kitagawa}, \citenamefont {Ishida}, \citenamefont {Nakamura}, \citenamefont
  {Shimizu}, \citenamefont {Homma}, \citenamefont {Li}, \citenamefont {Honda},\
  and\ \citenamefont {Aoki}}]{Tokunaga:2019}%
  \BibitemOpen
  \bibfield  {author} {\bibinfo {author} {\bibfnamefont {Y.}~\bibnamefont
  {Tokunaga}}, \bibinfo {author} {\bibfnamefont {H.}~\bibnamefont {Sakai}},
  \bibinfo {author} {\bibfnamefont {S.}~\bibnamefont {Kambe}}, \bibinfo
  {author} {\bibfnamefont {T.}~\bibnamefont {Hattori}}, \bibinfo {author}
  {\bibfnamefont {N.}~\bibnamefont {Higa}}, \bibinfo {author} {\bibfnamefont
  {G.}~\bibnamefont {Nakamine}}, \bibinfo {author} {\bibfnamefont
  {S.}~\bibnamefont {Kitagawa}}, \bibinfo {author} {\bibfnamefont
  {K.}~\bibnamefont {Ishida}}, \bibinfo {author} {\bibfnamefont
  {A.}~\bibnamefont {Nakamura}}, \bibinfo {author} {\bibfnamefont
  {Y.}~\bibnamefont {Shimizu}}, \bibinfo {author} {\bibfnamefont
  {Y.}~\bibnamefont {Homma}}, \bibinfo {author} {\bibfnamefont
  {D.}~\bibnamefont {Li}}, \bibinfo {author} {\bibfnamefont {F.}~\bibnamefont
  {Honda}}, \ and\ \bibinfo {author} {\bibfnamefont {D.}~\bibnamefont {Aoki}},\
  }\href {\doibase 10.7566/JPSJ.88.073701} {\bibfield  {journal} {\bibinfo
  {journal} {Journal of the Physical Society of Japan}\ }\textbf {\bibinfo
  {volume} {88}},\ \bibinfo {pages} {073701} (\bibinfo {year} {2019})},\
  \Eprint {http://arxiv.org/abs/https://doi.org/10.7566/JPSJ.88.073701}
  {https://doi.org/10.7566/JPSJ.88.073701} \BibitemShut {NoStop}%
\bibitem [{\citenamefont {Sundar}\ \emph {et~al.}(2019)\citenamefont {Sundar},
  \citenamefont {Gheidi}, \citenamefont {Akintola}, \citenamefont {Cote},
  \citenamefont {Dunsiger}, \citenamefont {Ran}, \citenamefont {Butch},
  \citenamefont {Saha}, \citenamefont {Paglione},\ and\ \citenamefont
  {Sonier}}]{Sundar:2019}%
  \BibitemOpen
  \bibfield  {author} {\bibinfo {author} {\bibfnamefont {S.}~\bibnamefont
  {Sundar}}, \bibinfo {author} {\bibfnamefont {S.}~\bibnamefont {Gheidi}},
  \bibinfo {author} {\bibfnamefont {K.}~\bibnamefont {Akintola}}, \bibinfo
  {author} {\bibfnamefont {A.~M.}\ \bibnamefont {Cote}}, \bibinfo {author}
  {\bibfnamefont {S.~R.}\ \bibnamefont {Dunsiger}}, \bibinfo {author}
  {\bibfnamefont {S.}~\bibnamefont {Ran}}, \bibinfo {author} {\bibfnamefont
  {N.~P.}\ \bibnamefont {Butch}}, \bibinfo {author} {\bibfnamefont {S.~R.}\
  \bibnamefont {Saha}}, \bibinfo {author} {\bibfnamefont {J.}~\bibnamefont
  {Paglione}}, \ and\ \bibinfo {author} {\bibfnamefont {J.~E.}\ \bibnamefont
  {Sonier}},\ }\href {\doibase {10.1103/PhysRevB.100.140502}} {\bibfield
  {journal} {\bibinfo  {journal} {{Phys. Rev. B}}\ }\textbf {\bibinfo {volume}
  {{100}}},\ \bibinfo {pages} {140502} (\bibinfo {year} {{2019}})}\BibitemShut
  {NoStop}%
\bibitem [{\citenamefont {Nakamine}\ \emph {et~al.}(2019)\citenamefont
  {Nakamine}, \citenamefont {Kitagawa}, \citenamefont {Ishida}, \citenamefont
  {Tokunaga}, \citenamefont {Sakai}, \citenamefont {Kambe}, \citenamefont
  {Nakamura}, \citenamefont {Shimizu}, \citenamefont {Homma}, \citenamefont
  {Li}, \citenamefont {Honda},\ and\ \citenamefont {Aoki}}]{Nakamine:2019}%
  \BibitemOpen
  \bibfield  {author} {\bibinfo {author} {\bibfnamefont {G.}~\bibnamefont
  {Nakamine}}, \bibinfo {author} {\bibfnamefont {S.}~\bibnamefont {Kitagawa}},
  \bibinfo {author} {\bibfnamefont {K.}~\bibnamefont {Ishida}}, \bibinfo
  {author} {\bibfnamefont {Y.}~\bibnamefont {Tokunaga}}, \bibinfo {author}
  {\bibfnamefont {H.}~\bibnamefont {Sakai}}, \bibinfo {author} {\bibfnamefont
  {S.}~\bibnamefont {Kambe}}, \bibinfo {author} {\bibfnamefont
  {A.}~\bibnamefont {Nakamura}}, \bibinfo {author} {\bibfnamefont
  {Y.}~\bibnamefont {Shimizu}}, \bibinfo {author} {\bibfnamefont
  {Y.}~\bibnamefont {Homma}}, \bibinfo {author} {\bibfnamefont
  {D.}~\bibnamefont {Li}}, \bibinfo {author} {\bibfnamefont {F.}~\bibnamefont
  {Honda}}, \ and\ \bibinfo {author} {\bibfnamefont {D.}~\bibnamefont {Aoki}},\
  }\href {\doibase {10.7566/JPSJ.88.113703}} {\bibfield  {journal} {\bibinfo
  {journal} {{Jour. Phys. Soc. Jpn.}}\ }\textbf {\bibinfo {volume} {{88}}},\
  \bibinfo {pages} {{113703}} (\bibinfo {year} {{2019}})}\BibitemShut {NoStop}%
\bibitem [{\citenamefont {Braithwaite}\ \emph {et~al.}(2019)\citenamefont
  {Braithwaite}, \citenamefont {Valiska}, \citenamefont {Knebel}, \citenamefont
  {Lapertot}, \citenamefont {Brison}, \citenamefont {Pourret}, \citenamefont
  {Zhitomirsky}, \citenamefont {Flouquet}, \citenamefont {Honda},\ and\
  \citenamefont {Aoki}}]{Braithwaite:2019}%
  \BibitemOpen
  \bibfield  {author} {\bibinfo {author} {\bibfnamefont {D.}~\bibnamefont
  {Braithwaite}}, \bibinfo {author} {\bibfnamefont {M.}~\bibnamefont
  {Valiska}}, \bibinfo {author} {\bibfnamefont {G.}~\bibnamefont {Knebel}},
  \bibinfo {author} {\bibfnamefont {G.}~\bibnamefont {Lapertot}}, \bibinfo
  {author} {\bibfnamefont {J.-P.}\ \bibnamefont {Brison}}, \bibinfo {author}
  {\bibfnamefont {A.}~\bibnamefont {Pourret}}, \bibinfo {author} {\bibfnamefont
  {M.~E.}\ \bibnamefont {Zhitomirsky}}, \bibinfo {author} {\bibfnamefont
  {J.}~\bibnamefont {Flouquet}}, \bibinfo {author} {\bibfnamefont
  {F.}~\bibnamefont {Honda}}, \ and\ \bibinfo {author} {\bibfnamefont
  {D.}~\bibnamefont {Aoki}},\ }\href {\doibase {10.1038/s42005-019-0248-z}}
  {\bibfield  {journal} {\bibinfo  {journal} {{Communications Physics}}\
  }\textbf {\bibinfo {volume} {{2}}},\ \bibinfo {pages} {147} (\bibinfo {year}
  {{2019}})}\BibitemShut {NoStop}%
\bibitem [{\citenamefont {Hayes}\ \emph {et~al.}(2020)\citenamefont {Hayes},
  \citenamefont {Wei}, \citenamefont {Metz}, \citenamefont {Zhang},
  \citenamefont {Eo}, \citenamefont {Ran}, \citenamefont {Saha}, \citenamefont
  {Collini}, \citenamefont {Butch}, \citenamefont {Agterberg}, \citenamefont
  {Kapitulnik},\ and\ \citenamefont {Paglione}}]{Hayes:2020}%
  \BibitemOpen
  \bibfield  {author} {\bibinfo {author} {\bibfnamefont {I.~M.}\ \bibnamefont
  {Hayes}}, \bibinfo {author} {\bibfnamefont {D.~S.}\ \bibnamefont {Wei}},
  \bibinfo {author} {\bibfnamefont {T.}~\bibnamefont {Metz}}, \bibinfo {author}
  {\bibfnamefont {J.}~\bibnamefont {Zhang}}, \bibinfo {author} {\bibfnamefont
  {Y.~S.}\ \bibnamefont {Eo}}, \bibinfo {author} {\bibfnamefont
  {S.}~\bibnamefont {Ran}}, \bibinfo {author} {\bibfnamefont {S.~R.}\
  \bibnamefont {Saha}}, \bibinfo {author} {\bibfnamefont {J.}~\bibnamefont
  {Collini}}, \bibinfo {author} {\bibfnamefont {N.~P.}\ \bibnamefont {Butch}},
  \bibinfo {author} {\bibfnamefont {D.~F.}\ \bibnamefont {Agterberg}}, \bibinfo
  {author} {\bibfnamefont {A.}~\bibnamefont {Kapitulnik}}, \ and\ \bibinfo
  {author} {\bibfnamefont {J.}~\bibnamefont {Paglione}},\ }\href@noop {} {\
  (\bibinfo {year} {2020})},\ \Eprint {http://arxiv.org/abs/2002.02539}
  {arXiv:2002.02539 [cond-mat.str-el]} \BibitemShut {NoStop}%
\bibitem [{\citenamefont {Thomas}\ \emph {et~al.}(2020)\citenamefont {Thomas},
  \citenamefont {Santos}, \citenamefont {Christensen}, \citenamefont {Asaba},
  \citenamefont {Ronning}, \citenamefont {Thompson}, \citenamefont {Bauer},
  \citenamefont {Fernandes}, \citenamefont {Fabbris},\ and\ \citenamefont
  {Rosa}}]{Thomas:2020}%
  \BibitemOpen
  \bibfield  {author} {\bibinfo {author} {\bibfnamefont {S.~M.}\ \bibnamefont
  {Thomas}}, \bibinfo {author} {\bibfnamefont {F.~B.}\ \bibnamefont {Santos}},
  \bibinfo {author} {\bibfnamefont {M.~H.}\ \bibnamefont {Christensen}},
  \bibinfo {author} {\bibfnamefont {T.}~\bibnamefont {Asaba}}, \bibinfo
  {author} {\bibfnamefont {F.}~\bibnamefont {Ronning}}, \bibinfo {author}
  {\bibfnamefont {J.~D.}\ \bibnamefont {Thompson}}, \bibinfo {author}
  {\bibfnamefont {E.~D.}\ \bibnamefont {Bauer}}, \bibinfo {author}
  {\bibfnamefont {R.~M.}\ \bibnamefont {Fernandes}}, \bibinfo {author}
  {\bibfnamefont {G.}~\bibnamefont {Fabbris}}, \ and\ \bibinfo {author}
  {\bibfnamefont {P.~F.~S.}\ \bibnamefont {Rosa}},\ }\href@noop {} {\enquote
  {\bibinfo {title} {{Evidence for a pressure-induced antiferromagnetic quantum
  critical point in intermediate valence UTe$_2$}},}\ } (\bibinfo {year}
  {2020}),\ \Eprint {http://arxiv.org/abs/2005.01659} {arXiv:2005.01659
  [cond-mat.str-el]} \BibitemShut {NoStop}%
\bibitem [{\citenamefont {Machida}(2020)}]{Machida:2020}%
  \BibitemOpen
  \bibfield  {author} {\bibinfo {author} {\bibfnamefont {K.}~\bibnamefont
  {Machida}},\ }\href {\doibase 10.7566/JPSJ.89.033702} {\bibfield  {journal}
  {\bibinfo  {journal} {Journal of the Physical Society of Japan}\ }\textbf
  {\bibinfo {volume} {89}},\ \bibinfo {pages} {033702} (\bibinfo {year}
  {2020})},\ \Eprint
  {http://arxiv.org/abs/https://doi.org/10.7566/JPSJ.89.033702}
  {https://doi.org/10.7566/JPSJ.89.033702} \BibitemShut {NoStop}%
\bibitem [{\citenamefont {Kittaka}\ \emph {et~al.}(2020)\citenamefont
  {Kittaka}, \citenamefont {Shimizu}, \citenamefont {Sakakibara}, \citenamefont
  {Nakamura}, \citenamefont {Li}, \citenamefont {Homma}, \citenamefont {Honda},
  \citenamefont {Aoki},\ and\ \citenamefont {Machida}}]{Kittaka:2020}%
  \BibitemOpen
  \bibfield  {author} {\bibinfo {author} {\bibfnamefont {S.}~\bibnamefont
  {Kittaka}}, \bibinfo {author} {\bibfnamefont {Y.}~\bibnamefont {Shimizu}},
  \bibinfo {author} {\bibfnamefont {T.}~\bibnamefont {Sakakibara}}, \bibinfo
  {author} {\bibfnamefont {A.}~\bibnamefont {Nakamura}}, \bibinfo {author}
  {\bibfnamefont {D.}~\bibnamefont {Li}}, \bibinfo {author} {\bibfnamefont
  {Y.}~\bibnamefont {Homma}}, \bibinfo {author} {\bibfnamefont
  {F.}~\bibnamefont {Honda}}, \bibinfo {author} {\bibfnamefont
  {D.}~\bibnamefont {Aoki}}, \ and\ \bibinfo {author} {\bibfnamefont
  {K.}~\bibnamefont {Machida}},\ }\href {\doibase
  10.1103/PhysRevResearch.2.032014} {\bibfield  {journal} {\bibinfo  {journal}
  {Phys. Rev. Research}\ }\textbf {\bibinfo {volume} {2}},\ \bibinfo {pages}
  {032014} (\bibinfo {year} {2020})}\BibitemShut {NoStop}%
\bibitem [{\citenamefont {Bae}\ \emph {et~al.}(2019)\citenamefont {Bae},
  \citenamefont {Kim}, \citenamefont {Ran}, \citenamefont {Eo}, \citenamefont
  {Liu}, \citenamefont {Fuhrman}, \citenamefont {Paglione}, \citenamefont
  {Butch},\ and\ \citenamefont {Anlage}}]{Bae:2019}%
  \BibitemOpen
  \bibfield  {author} {\bibinfo {author} {\bibfnamefont {S.}~\bibnamefont
  {Bae}}, \bibinfo {author} {\bibfnamefont {H.}~\bibnamefont {Kim}}, \bibinfo
  {author} {\bibfnamefont {S.}~\bibnamefont {Ran}}, \bibinfo {author}
  {\bibfnamefont {Y.~S.}\ \bibnamefont {Eo}}, \bibinfo {author} {\bibfnamefont
  {I.-L.}\ \bibnamefont {Liu}}, \bibinfo {author} {\bibfnamefont
  {W.}~\bibnamefont {Fuhrman}}, \bibinfo {author} {\bibfnamefont
  {J.}~\bibnamefont {Paglione}}, \bibinfo {author} {\bibfnamefont {N.~P.}\
  \bibnamefont {Butch}}, \ and\ \bibinfo {author} {\bibfnamefont
  {S.}~\bibnamefont {Anlage}},\ }\href@noop {} {\enquote {\bibinfo {title}
  {Anomalous normal fluid response in a chiral superconductor},}\ } (\bibinfo
  {year} {2019}),\ \Eprint {http://arxiv.org/abs/1909.09032} {arXiv:1909.09032
  [cond-mat.supr-con]} \BibitemShut {NoStop}%
\bibitem [{\citenamefont {Jiao}\ \emph {et~al.}(2020)\citenamefont {Jiao},
  \citenamefont {Howard}, \citenamefont {Ran}, \citenamefont {Wang},
  \citenamefont {Rodriguez}, \citenamefont {Sigrist}, \citenamefont {Wang},
  \citenamefont {Butch},\ and\ \citenamefont {Madhavan}}]{Jiao:2020}%
  \BibitemOpen
  \bibfield  {author} {\bibinfo {author} {\bibfnamefont {L.}~\bibnamefont
  {Jiao}}, \bibinfo {author} {\bibfnamefont {S.}~\bibnamefont {Howard}},
  \bibinfo {author} {\bibfnamefont {S.}~\bibnamefont {Ran}}, \bibinfo {author}
  {\bibfnamefont {Z.}~\bibnamefont {Wang}}, \bibinfo {author} {\bibfnamefont
  {J.~O.}\ \bibnamefont {Rodriguez}}, \bibinfo {author} {\bibfnamefont
  {M.}~\bibnamefont {Sigrist}}, \bibinfo {author} {\bibfnamefont
  {Z.}~\bibnamefont {Wang}}, \bibinfo {author} {\bibfnamefont {N.~P.}\
  \bibnamefont {Butch}}, \ and\ \bibinfo {author} {\bibfnamefont
  {V.}~\bibnamefont {Madhavan}},\ }\href {\doibase 10.1038/s41586-020-2122-2}
  {\bibfield  {journal} {\bibinfo  {journal} {Nature}\ }\textbf {\bibinfo
  {volume} {579}},\ \bibinfo {pages} {523} (\bibinfo {year}
  {2020})}\BibitemShut {NoStop}%
\bibitem [{\citenamefont {Aoki}\ \emph
  {et~al.}(2019{\natexlab{b}})\citenamefont {Aoki}, \citenamefont {Ishida},\
  and\ \citenamefont {Flouquet}}]{Aoki:2019}%
  \BibitemOpen
  \bibfield  {author} {\bibinfo {author} {\bibfnamefont {D.}~\bibnamefont
  {Aoki}}, \bibinfo {author} {\bibfnamefont {K.}~\bibnamefont {Ishida}}, \ and\
  \bibinfo {author} {\bibfnamefont {J.}~\bibnamefont {Flouquet}},\ }\href
  {\doibase {10.7566/JPSJ.88.022001}} {\bibfield  {journal} {\bibinfo
  {journal} {{J. Phys. Soc. Jpn.}}\ }\textbf {\bibinfo {volume} {{88}}},\
  \bibinfo {pages} {022001} (\bibinfo {year}
  {{2019}}{\natexlab{b}})}\BibitemShut {NoStop}%
\bibitem [{\citenamefont {Shick}\ and\ \citenamefont
  {Pickett}(2019)}]{Shick:2019}%
  \BibitemOpen
  \bibfield  {author} {\bibinfo {author} {\bibfnamefont {A.~B.}\ \bibnamefont
  {Shick}}\ and\ \bibinfo {author} {\bibfnamefont {W.~E.}\ \bibnamefont
  {Pickett}},\ }\href {\doibase {10.1103/PhysRevB.100.134502}} {\bibfield
  {journal} {\bibinfo  {journal} {{Phys. Rev. B}}\ }\textbf {\bibinfo {volume}
  {{100}}},\ \bibinfo {pages} {{134502}} (\bibinfo {year}
  {{2019}})}\BibitemShut {NoStop}%
\bibitem [{\citenamefont {Xu}\ \emph {et~al.}(2019)\citenamefont {Xu},
  \citenamefont {Sheng},\ and\ \citenamefont {Yang}}]{Xu:2019}%
  \BibitemOpen
  \bibfield  {author} {\bibinfo {author} {\bibfnamefont {Y.}~\bibnamefont
  {Xu}}, \bibinfo {author} {\bibfnamefont {Y.}~\bibnamefont {Sheng}}, \ and\
  \bibinfo {author} {\bibfnamefont {Y.-F.}\ \bibnamefont {Yang}},\ }\href
  {\doibase {10.1103/PhysRevLett.123.217002}} {\bibfield  {journal} {\bibinfo
  {journal} {{Phys. Rev. Lett.}}\ }\textbf {\bibinfo {volume} {{123}}},\
  \bibinfo {pages} {217002} (\bibinfo {year} {{2019}})}\BibitemShut {NoStop}%
\bibitem [{\citenamefont {Ishizuka}\ \emph {et~al.}(2019)\citenamefont
  {Ishizuka}, \citenamefont {Sumita}, \citenamefont {Daido},\ and\
  \citenamefont {Yanase}}]{Ishizuka:2019}%
  \BibitemOpen
  \bibfield  {author} {\bibinfo {author} {\bibfnamefont {J.}~\bibnamefont
  {Ishizuka}}, \bibinfo {author} {\bibfnamefont {S.}~\bibnamefont {Sumita}},
  \bibinfo {author} {\bibfnamefont {A.}~\bibnamefont {Daido}}, \ and\ \bibinfo
  {author} {\bibfnamefont {Y.}~\bibnamefont {Yanase}},\ }\href {\doibase
  {10.1103/PhysRevLett.123.217001}} {\bibfield  {journal} {\bibinfo  {journal}
  {{Phys. Rev. Lett.}}\ }\textbf {\bibinfo {volume} {{123}}},\ \bibinfo {pages}
  {{217001}} (\bibinfo {year} {{2019}})}\BibitemShut {NoStop}%
\bibitem [{\citenamefont {Metz}\ \emph {et~al.}(2019)\citenamefont {Metz},
  \citenamefont {Bae}, \citenamefont {Ran}, \citenamefont {Liu}, \citenamefont
  {Eo}, \citenamefont {Fuhrman}, \citenamefont {Agterberg}, \citenamefont
  {Anlage}, \citenamefont {Butch},\ and\ \citenamefont {Paglione}}]{Metz:2019}%
  \BibitemOpen
  \bibfield  {author} {\bibinfo {author} {\bibfnamefont {T.}~\bibnamefont
  {Metz}}, \bibinfo {author} {\bibfnamefont {S.}~\bibnamefont {Bae}}, \bibinfo
  {author} {\bibfnamefont {S.}~\bibnamefont {Ran}}, \bibinfo {author}
  {\bibfnamefont {I.-L.}\ \bibnamefont {Liu}}, \bibinfo {author} {\bibfnamefont
  {Y.~S.}\ \bibnamefont {Eo}}, \bibinfo {author} {\bibfnamefont {W.~T.}\
  \bibnamefont {Fuhrman}}, \bibinfo {author} {\bibfnamefont {D.~F.}\
  \bibnamefont {Agterberg}}, \bibinfo {author} {\bibfnamefont {S.}~\bibnamefont
  {Anlage}}, \bibinfo {author} {\bibfnamefont {N.~P.}\ \bibnamefont {Butch}}, \
  and\ \bibinfo {author} {\bibfnamefont {J.}~\bibnamefont {Paglione}},\ }\href
  {\doibase {10.1103/PhysRevB.100.220504}} {\bibfield  {journal} {\bibinfo
  {journal} {{Phys. Rev. B}}\ }\textbf {\bibinfo {volume} {{100}}},\ \bibinfo
  {pages} {{220504}} (\bibinfo {year} {{2019}})}\BibitemShut {NoStop}%
\bibitem [{\citenamefont {Fujimori}\ \emph {et~al.}(2019)\citenamefont
  {Fujimori}, \citenamefont {Kawasaki}, \citenamefont {Takeda}, \citenamefont
  {Yamagami}, \citenamefont {Nakamura}, \citenamefont {Homma},\ and\
  \citenamefont {Aoki}}]{Fujimori:2019}%
  \BibitemOpen
  \bibfield  {author} {\bibinfo {author} {\bibfnamefont {S.-i.}\ \bibnamefont
  {Fujimori}}, \bibinfo {author} {\bibfnamefont {I.}~\bibnamefont {Kawasaki}},
  \bibinfo {author} {\bibfnamefont {Y.}~\bibnamefont {Takeda}}, \bibinfo
  {author} {\bibfnamefont {H.}~\bibnamefont {Yamagami}}, \bibinfo {author}
  {\bibfnamefont {A.}~\bibnamefont {Nakamura}}, \bibinfo {author}
  {\bibfnamefont {Y.}~\bibnamefont {Homma}}, \ and\ \bibinfo {author}
  {\bibfnamefont {D.}~\bibnamefont {Aoki}},\ }\href {\doibase
  10.7566/JPSJ.88.103701} {\bibfield  {journal} {\bibinfo  {journal} {Journal
  of the Physical Society of Japan}\ }\textbf {\bibinfo {volume} {88}},\
  \bibinfo {pages} {103701} (\bibinfo {year} {2019})},\ \Eprint
  {http://arxiv.org/abs/https://doi.org/10.7566/JPSJ.88.103701}
  {https://doi.org/10.7566/JPSJ.88.103701} \BibitemShut {NoStop}%
\bibitem [{\citenamefont {Miao}\ \emph {et~al.}(2020)\citenamefont {Miao},
  \citenamefont {Liu}, \citenamefont {Xu}, \citenamefont {Kotta}, \citenamefont
  {Kang}, \citenamefont {Ran}, \citenamefont {Paglione}, \citenamefont
  {Kotliar}, \citenamefont {Butch}, \citenamefont {Denlinger},\ and\
  \citenamefont {Wray}}]{Miao:2020}%
  \BibitemOpen
  \bibfield  {author} {\bibinfo {author} {\bibfnamefont {L.}~\bibnamefont
  {Miao}}, \bibinfo {author} {\bibfnamefont {S.}~\bibnamefont {Liu}}, \bibinfo
  {author} {\bibfnamefont {Y.}~\bibnamefont {Xu}}, \bibinfo {author}
  {\bibfnamefont {E.~C.}\ \bibnamefont {Kotta}}, \bibinfo {author}
  {\bibfnamefont {C.-J.}\ \bibnamefont {Kang}}, \bibinfo {author}
  {\bibfnamefont {S.}~\bibnamefont {Ran}}, \bibinfo {author} {\bibfnamefont
  {J.}~\bibnamefont {Paglione}}, \bibinfo {author} {\bibfnamefont
  {G.}~\bibnamefont {Kotliar}}, \bibinfo {author} {\bibfnamefont {N.~P.}\
  \bibnamefont {Butch}}, \bibinfo {author} {\bibfnamefont {J.~D.}\ \bibnamefont
  {Denlinger}}, \ and\ \bibinfo {author} {\bibfnamefont {L.~A.}\ \bibnamefont
  {Wray}},\ }\href {\doibase {10.1103/PhysRevLett.124.076401}} {\bibfield
  {journal} {\bibinfo  {journal} {{Phys. Rev. Lett.}}\ }\textbf {\bibinfo
  {volume} {{124}}},\ \bibinfo {pages} {{076401}} (\bibinfo {year}
  {{2020}})}\BibitemShut {NoStop}%
\bibitem [{\citenamefont {Weinert}\ \emph {et~al.}(2009)\citenamefont
  {Weinert}, \citenamefont {Schneider}, \citenamefont {Podloucky},\ and\
  \citenamefont {Redinger}}]{Mike_FLAPW_2009}%
  \BibitemOpen
  \bibfield  {author} {\bibinfo {author} {\bibfnamefont {M.}~\bibnamefont
  {Weinert}}, \bibinfo {author} {\bibfnamefont {G.}~\bibnamefont {Schneider}},
  \bibinfo {author} {\bibfnamefont {R.}~\bibnamefont {Podloucky}}, \ and\
  \bibinfo {author} {\bibfnamefont {J.}~\bibnamefont {Redinger}},\ }\href
  {\doibase 10.1088/0953-8984/21/8/084201} {\bibfield  {journal} {\bibinfo
  {journal} {J. Phys. Condens. Matter}\ }\textbf {\bibinfo {volume} {21}},\
  \bibinfo {pages} {084201} (\bibinfo {year} {2009})}\BibitemShut {NoStop}%
\bibitem [{\citenamefont {Fu}\ and\ \citenamefont {Kane}(2007)}]{FuKane:2007}%
  \BibitemOpen
  \bibfield  {author} {\bibinfo {author} {\bibfnamefont {L.}~\bibnamefont
  {Fu}}\ and\ \bibinfo {author} {\bibfnamefont {C.~L.}\ \bibnamefont {Kane}},\
  }\href {\doibase 10.1103/PhysRevB.76.045302} {\bibfield  {journal} {\bibinfo
  {journal} {Phys. Rev. B}\ }\textbf {\bibinfo {volume} {76}},\ \bibinfo
  {pages} {045302} (\bibinfo {year} {2007})}\BibitemShut {NoStop}%
\bibitem [{\citenamefont {Ramires}\ and\ \citenamefont
  {Sigrist}(2016)}]{Ramires:2016}%
  \BibitemOpen
  \bibfield  {author} {\bibinfo {author} {\bibfnamefont {A.}~\bibnamefont
  {Ramires}}\ and\ \bibinfo {author} {\bibfnamefont {M.}~\bibnamefont
  {Sigrist}},\ }\href@noop {} {\bibfield  {journal} {\bibinfo  {journal} {Phys.
  Rev. B}\ }\textbf {\bibinfo {volume} {94}},\ \bibinfo {pages} {104501}
  (\bibinfo {year} {2016})}\BibitemShut {NoStop}%
\bibitem [{\citenamefont {Ramires}\ \emph {et~al.}(2018)\citenamefont
  {Ramires}, \citenamefont {Agterberg},\ and\ \citenamefont
  {Sigrist}}]{Ramires:2018}%
  \BibitemOpen
  \bibfield  {author} {\bibinfo {author} {\bibfnamefont {A.}~\bibnamefont
  {Ramires}}, \bibinfo {author} {\bibfnamefont {D.~F.}\ \bibnamefont
  {Agterberg}}, \ and\ \bibinfo {author} {\bibfnamefont {M.}~\bibnamefont
  {Sigrist}},\ }\href@noop {} {\bibfield  {journal} {\bibinfo  {journal} {Phys.
  Rev. B}\ }\textbf {\bibinfo {volume} {98}},\ \bibinfo {pages} {024501}
  (\bibinfo {year} {2018})}\BibitemShut {NoStop}%
\bibitem [{\citenamefont {Yarzhemsky}\ and\ \citenamefont
  {Teplyakov}(2020)}]{Yarzhemsky:2020}%
  \BibitemOpen
  \bibfield  {author} {\bibinfo {author} {\bibfnamefont {V.~G.}\ \bibnamefont
  {Yarzhemsky}}\ and\ \bibinfo {author} {\bibfnamefont {E.~A.}\ \bibnamefont
  {Teplyakov}},\ }\href@noop {} {\enquote {\bibinfo {title} {{Time-reversal
  symmetry and the structure of superconducting order Parameter of nearly
  ferromagnetic spin-triplet superconductor UTe$_2$}},}\ } (\bibinfo {year}
  {2020}),\ \Eprint {http://arxiv.org/abs/2001.02963} {arXiv:2001.02963
  [cond-mat.supr-con]} \BibitemShut {NoStop}%
\bibitem [{\citenamefont {Nevidomskyy}(2020)}]{Nevidomskyy:2020}%
  \BibitemOpen
  \bibfield  {author} {\bibinfo {author} {\bibfnamefont {A.~H.}\ \bibnamefont
  {Nevidomskyy}},\ }\href@noop {} {\enquote {\bibinfo {title} {{Stability of a
  nonunitary triplet pairing on the border of magnetism in UTe$_2$}},}\ }
  (\bibinfo {year} {2020}),\ \Eprint {http://arxiv.org/abs/2001.02699}
  {arXiv:2001.02699 [cond-mat.supr-con]} \BibitemShut {NoStop}%
\bibitem [{\citenamefont {Aoki}\ \emph {et~al.}(2020)\citenamefont {Aoki},
  \citenamefont {Honda}, \citenamefont {Knebel}, \citenamefont {Braithwaite},
  \citenamefont {Nakamura}, \citenamefont {Li}, \citenamefont {Homma},
  \citenamefont {Shimizu}, \citenamefont {Sato}, \citenamefont {Brison},\ and\
  \citenamefont {Flouquet}}]{Aoki:2020}%
  \BibitemOpen
  \bibfield  {author} {\bibinfo {author} {\bibfnamefont {D.}~\bibnamefont
  {Aoki}}, \bibinfo {author} {\bibfnamefont {F.}~\bibnamefont {Honda}},
  \bibinfo {author} {\bibfnamefont {G.}~\bibnamefont {Knebel}}, \bibinfo
  {author} {\bibfnamefont {D.}~\bibnamefont {Braithwaite}}, \bibinfo {author}
  {\bibfnamefont {A.}~\bibnamefont {Nakamura}}, \bibinfo {author}
  {\bibfnamefont {D.}~\bibnamefont {Li}}, \bibinfo {author} {\bibfnamefont
  {Y.}~\bibnamefont {Homma}}, \bibinfo {author} {\bibfnamefont
  {Y.}~\bibnamefont {Shimizu}}, \bibinfo {author} {\bibfnamefont {Y.~J.}\
  \bibnamefont {Sato}}, \bibinfo {author} {\bibfnamefont {J.-P.}\ \bibnamefont
  {Brison}}, \ and\ \bibinfo {author} {\bibfnamefont {J.}~\bibnamefont
  {Flouquet}},\ }\href {\doibase 10.7566/JPSJ.89.053705} {\bibfield  {journal}
  {\bibinfo  {journal} {Journal of the Physical Society of Japan}\ }\textbf
  {\bibinfo {volume} {89}},\ \bibinfo {pages} {053705} (\bibinfo {year}
  {2020})},\ \Eprint
  {http://arxiv.org/abs/https://doi.org/10.7566/JPSJ.89.053705}
  {https://doi.org/10.7566/JPSJ.89.053705} \BibitemShut {NoStop}%
\bibitem [{\citenamefont {Lin}\ \emph {et~al.}(2020)\citenamefont {Lin},
  \citenamefont {Campbell}, \citenamefont {Ran}, \citenamefont {Liu},
  \citenamefont {Kim}, \citenamefont {Nevidomskyy}, \citenamefont {Graf},
  \citenamefont {Butch},\ and\ \citenamefont {Paglione}}]{Lin:2020}%
  \BibitemOpen
  \bibfield  {author} {\bibinfo {author} {\bibfnamefont {W.-C.}\ \bibnamefont
  {Lin}}, \bibinfo {author} {\bibfnamefont {D.~J.}\ \bibnamefont {Campbell}},
  \bibinfo {author} {\bibfnamefont {S.}~\bibnamefont {Ran}}, \bibinfo {author}
  {\bibfnamefont {I.-L.}\ \bibnamefont {Liu}}, \bibinfo {author} {\bibfnamefont
  {H.}~\bibnamefont {Kim}}, \bibinfo {author} {\bibfnamefont {A.~H.}\
  \bibnamefont {Nevidomskyy}}, \bibinfo {author} {\bibfnamefont
  {D.}~\bibnamefont {Graf}}, \bibinfo {author} {\bibfnamefont {N.~P.}\
  \bibnamefont {Butch}}, \ and\ \bibinfo {author} {\bibfnamefont
  {J.}~\bibnamefont {Paglione}},\ }\href@noop {} {\enquote {\bibinfo {title}
  {Tuning magnetic confinement of spin-triplet superconductivity},}\ }
  (\bibinfo {year} {2020}),\ \Eprint {http://arxiv.org/abs/2002.12885}
  {arXiv:2002.12885 [cond-mat.supr-con]} \BibitemShut {NoStop}%
\bibitem [{\citenamefont {Lange}\ and\ \citenamefont
  {Kotliar}(1999)}]{Lange:1999}%
  \BibitemOpen
  \bibfield  {author} {\bibinfo {author} {\bibfnamefont {E.}~\bibnamefont
  {Lange}}\ and\ \bibinfo {author} {\bibfnamefont {G.}~\bibnamefont
  {Kotliar}},\ }\href {\doibase 10.1103/PhysRevLett.82.1317} {\bibfield
  {journal} {\bibinfo  {journal} {Phys. Rev. Lett.}\ }\textbf {\bibinfo
  {volume} {82}},\ \bibinfo {pages} {1317} (\bibinfo {year}
  {1999})}\BibitemShut {NoStop}%
\bibitem [{\citenamefont {Brydon}\ \emph {et~al.}(2018)\citenamefont {Brydon},
  \citenamefont {Agterberg}, \citenamefont {Menke},\ and\ \citenamefont
  {Timm}}]{Brydon:2018}%
  \BibitemOpen
  \bibfield  {author} {\bibinfo {author} {\bibfnamefont {P.~M.~R.}\
  \bibnamefont {Brydon}}, \bibinfo {author} {\bibfnamefont {D.~F.}\
  \bibnamefont {Agterberg}}, \bibinfo {author} {\bibfnamefont {H.}~\bibnamefont
  {Menke}}, \ and\ \bibinfo {author} {\bibfnamefont {C.}~\bibnamefont {Timm}},\
  }\href {\doibase 10.1103/PhysRevB.98.224509} {\bibfield  {journal} {\bibinfo
  {journal} {Phys. Rev. B}\ }\textbf {\bibinfo {volume} {98}},\ \bibinfo
  {pages} {224509} (\bibinfo {year} {2018})}\BibitemShut {NoStop}%
\bibitem [{\citenamefont {Brydon}\ \emph {et~al.}(2019)\citenamefont {Brydon},
  \citenamefont {Abergel}, \citenamefont {Agterberg},\ and\ \citenamefont
  {Yakovenko}}]{Brydon:2019}%
  \BibitemOpen
  \bibfield  {author} {\bibinfo {author} {\bibfnamefont {P.~M.~R.}\
  \bibnamefont {Brydon}}, \bibinfo {author} {\bibfnamefont {D.~S.~L.}\
  \bibnamefont {Abergel}}, \bibinfo {author} {\bibfnamefont {D.~F.}\
  \bibnamefont {Agterberg}}, \ and\ \bibinfo {author} {\bibfnamefont {V.~M.}\
  \bibnamefont {Yakovenko}},\ }\href {\doibase 10.1103/PhysRevX.9.031025}
  {\bibfield  {journal} {\bibinfo  {journal} {Phys. Rev. X}\ }\textbf {\bibinfo
  {volume} {9}},\ \bibinfo {pages} {031025} (\bibinfo {year}
  {2019})}\BibitemShut {NoStop}%
\bibitem [{\citenamefont {Taylor}\ and\ \citenamefont
  {Kallin}(2012)}]{Taylor:2012}%
  \BibitemOpen
  \bibfield  {author} {\bibinfo {author} {\bibfnamefont {E.}~\bibnamefont
  {Taylor}}\ and\ \bibinfo {author} {\bibfnamefont {C.}~\bibnamefont
  {Kallin}},\ }\href {\doibase 10.1103/PhysRevLett.108.157001} {\bibfield
  {journal} {\bibinfo  {journal} {Phys. Rev. Lett.}\ }\textbf {\bibinfo
  {volume} {108}},\ \bibinfo {pages} {157001} (\bibinfo {year}
  {2012})}\BibitemShut {NoStop}%
\end{thebibliography}%


\begin{thebibliography}{9}
\bibitem{ran19} S. Ran {\it et al}., Science {\bf 365}, 684 (2019).
\bibitem{shi19} A.B. Shick and W.E. Pickett, Phys. Rev. B {\bf 100}, 134502 (2019).
\bibitem{ish19} Jun Ishizuka, Shuntaro Sumita, Akito Daido, and Youichi Yanase, Phys. Rev. Lett. {\bf 123}, 217001 (2019).
\bibitem{xu19} Yuanji Xu, Yutao Sheng, and Yi-feng Yangz
Phys. Rev. Lett. {\bf 123}, 217002 (2019).
\bibitem{met19} T. Metz {\it et al}., Phys. Rev. B, in press (2019).
\bibitem{bra19} D. Braithwaite, M. Vali\UTF{0161}ka, G. Knebel, G. Lapertot, J.-P. Brison, A. Pourret, M. E. Zhitomirsky, J. Flouquet, F. Honda, and  D. Aoki, Nature Communications Physics {\bf 2}, 147 (2019).
\bibitem{jia19} Jiao {\it at al}., arXiv:1908.02846 (2019). 
\bibitem{koz16} V. Kozii, J.W.F. Venderbos, and L. Fu, Science Advances {\bf 2}, e1601835 (2016).





\end{thebibliography}

\end{document}